\documentclass[aps, secnumarabic,twocolumn,nobalancelastpage,amsmath,amssymb,nofootinbib]{revtex4}

\usepackage{graphicx}
\usepackage{dcolumn}
\usepackage{bm}
\usepackage{ulem}
\usepackage{amssymb}
\usepackage{braket}
\usepackage{verbatim}
\usepackage[capitalise]{cleveref} 
\usepackage{hyperref}

\newcommand{\op}[1]{\hat{#1}}								
\newcommand{\dg}{^\dagger}

\newcommand{\ketbra}[2]{\ket{#1}\!\bra{#2}}				

\begin{document}


\title{Simple, smooth and fast pulses for dispersive measurements in cavities and quantum networks}

\author{Felix Motzoi}
\affiliation{Department of Physics and Astronomy, Aarhus University, DK-8000 Aarhus C, Denmark}
\author{Christian Dickel}
\affiliation{Kavli Institute of Nanoscience, Delft University of Technology, Lorentzweg 1, 2628 CJ Delft, The Netherlands}
\author{Lukas Buchmann}
\affiliation{Department of Physics and Astronomy, Aarhus University, DK-8000 Aarhus C, Denmark}

\date{\today}

\begin{abstract}
We demonstrate a dispersive measurement pulse shaping technique that allows for arbitrarily fast quantum non-demolition, single-quadrature measurements of non-linear systems and unconditionally leaves the measurement resonator empty. For single-qubit measurements, current measurements are limited to the 99\% fidelity range due to relaxation during the process. However, trying to go to shorter times to circumvent this with square or composite digital pulses leads to leftover cavity population after measurement of the same order of error.  These effects can be suppressed using simple smooth pulse shapes from a similar family of pulses as DRAG shaping, used in the context of leakage removal in superconducting qubits; here, it can be derived exactly for arbitrarily many measured modes.  Beyond single qubits, the measurement pulses are fully general to dispersive measurement systems. This includes multi-qubit and multi-state (leakage) measurements where the measurement can be done in a single shot and with a single homodyne phase.  Another major challenge for fast measurement is depopulating Purcell filter cavities, which we show can readily be achieved using derivative shaping. Finally, we show how to apply the technique to cascaded cavity systems, e.g~for fast remote entanglement generation. 
\end{abstract}

\maketitle
\section{introduction}

The fast and efficient determination of the resonance frequency of a resonator with known linewidth is a crucial task in the context of quantum technology. 

When a cavity is coupled quantum system (e.g.~qubit) while far detuned in energy, i.e.~dispersively couple, the qubit state can be inferred from the cavity frequency.
This dispersive qubit readout has been widely used to implement high-fidelity quantum non-demolition (QND) measurements~\cite{Braginsky80},
particularly in circuit QED \cite{Blais04}.

High-fidelity QND measurements are essential to qubit initialization by measurement~\cite{Riste12} and ancilla-based syndrome measurements for quantum error correction~\cite{Kelly15,Cramer16}.
However, measurement has typically lagged in performance compared with other constituent processes such as single-qubit and two-qubit gates, which have reached $10^{-3}$ in several implementations \cite{Barends:2014th,Benhelm:2008fy}.
The measurements also presently dominate the clock-cycle duration in quantum error correction schemes~\cite{Kelly15} and hence are the leading cause of qubit decoherence.
The main impediment is that open-system control methods need to face a trade off between the retrieval of information and the unwanted leakage of information into the environment through the same output ports. 

\begin{figure}	
\includegraphics[width=.45\textwidth]{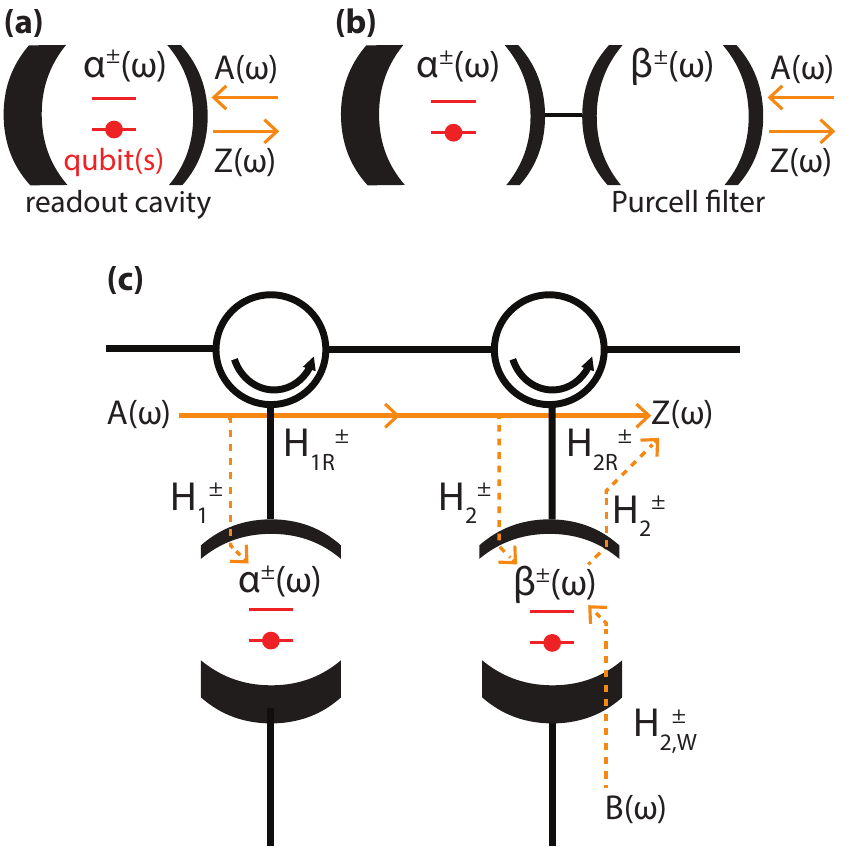}
\caption{Example topologies for dispersive measurement in quantum networks. (A) Standard (multi-)qubit/qutrit measurement by measuring in reflection on the readout cavity. (B) Purcell filter, where a second cavity is used to suppress qubit relaxation into the environment. (C) Sequential cavity readout scheme for generating entanglement}\label{fig:setup}
\end{figure}

Several attempts have been made to use digital pulse sequences \cite{McClure16,Bultink17}, rather than allowing the cavity to empty out naturally.
Applying such schemes beyond the case of a single qubit has not been achieved, as the run-time of numerical optimization, and the speed and accuracy of experimental implementation all scale poorly with the number of qubits. 
Even for single qubits, requiring a steady-state of the radiation field together with digital ramp-up and ramp-down pulses result in overall longer measurements, only reducing times by half compared to square pulses. Additionally, the easier-to-optimize digital pulse sequences are heavily filtered before they reach the low-temperature cavity and are poor matches for experiment.
Altogether, these factors have limited experimental fidelities to the 99\% range, even as quantum efficiencies have nearly reached the quantum-limited regime using parametric amplifiers \cite{Siddiqi2004}.

Although measurements are often idealized as projective and/or evolving under a constant operator, the requirement to fall well below fault-tolerance thresholds (and thus greatly reduce system sizes) invariably requires modeling fast time dynamics, without a steady state.

The key metrics for an ideal QND measurement protocol are speed, contrast, and reproducibility.
A fast, high-efficiency measurement ensures that the desired information is retrieved without being left behind in the measurement apparatus or lost to the environment; contrast between distinguishable quantum states assures the accuracy of the measurement; reproducibility assures us that future operations will not suffer from the present measurement. 
A standard way to optimize fast time dynamics is using numerical optimal control theory \cite{Egger2014}, however this requires pulses tailored to individual experimental realizations \cite{Motzoi2011}, where such fine tuning leads to solutions sensitive to experimental imperfections. 

On the other hand, a growing body of work has focused on analytical solutions that artificially enforce adiabaticity criteria on the dynamics via extra control fields \cite{shortcuts-1,Demirplak_2003,Demirplak_2008,Motzoi2009}.  To date, such solutions have either controlled a single diabatic gap energy or involved perturbative approximate solutions with decreased speed. Another problematic aspect has been applying such ``shortcuts" to open quantum systems where the concept of a gap energy is more evasive.

Here, we show how to incorporate both these objectives in a resonator assisted quantum-state readout protocol with a simple, experimentally robust solution. Our proposal bypasses the quantum speed limit imposed by the communication rate with the environment by solving exactly a system of artificial adiabaticity conditions for a number of different gapped Lorentzians corresponding to different probed states. Our solutions significantly outperform earlier protocols by providing exact and closed-form solutions to continuous and filtered measurement dynamics.

\begin{figure}	
	\includegraphics[width=.47\textwidth]{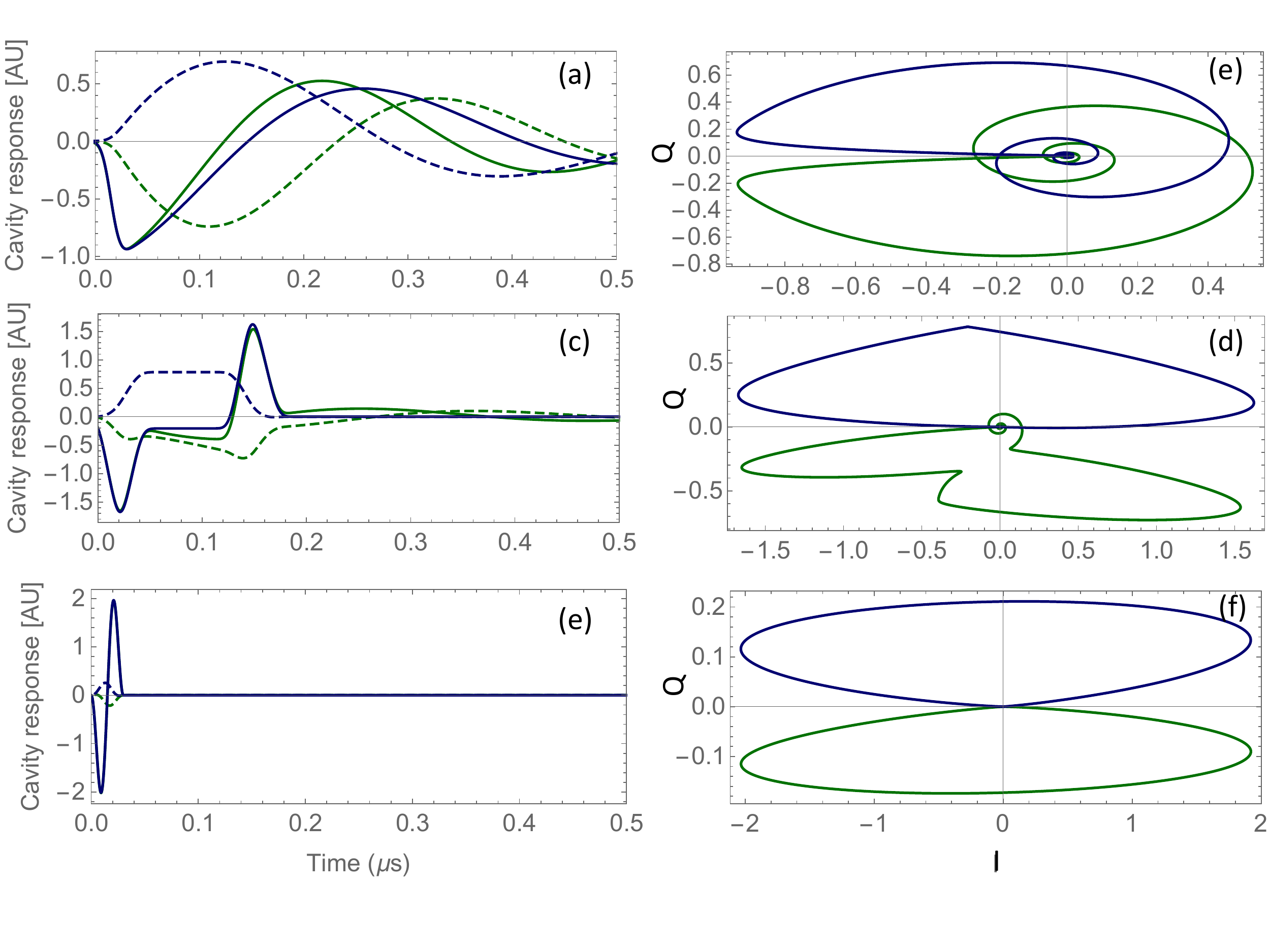}

\caption{Evolution of a cavity dispersively coupled to one qubit for three different measurement pulse sequences: square pulse (top row); numerically-optimized CLEAR-type digital pulse sequence (middle row); and Gaussian with derivative corrections (bottom row). Solid and dashed lines correspond to time-resolved values of I and Q quadratures for state 0 (green) and 1(blue) in (a), (c) and (e). Corresponding evolution in the I-Q plane is shown in (b), (d) and (f). Note that in the bottom plots the solid I quadrature lines are perfectly overlapping, indicating homodyne measurement. Simulated parameters are $\kappa=2$MHz, $\Delta=0.3$MHz, $\chi=2$MHz.}  \label{fig:1Q}
\end{figure}

\section{Fast dispersive measurement}

We model the effect of coherent measurement fields on a quantum network \cite{Kimble:2008dk,Duan2010}.  In general terms there exists a broad array of such systems, ranging from single-cavity dispersive measurements (used in circuit QED), to sophisticated optical networks used for Boson sampling problems, to quantum circuitry used for error correction and entangled photon or matter state generation. Our focus is on the dynamics of time-resolved input fields ($A(t)$) and in particular the measured signal $Z(t)$.

Let a quantum network $\mathcal{Q}$, consisting of passive linear elements, be coupled dispersively to a set of few-level quantum systems. 
 Given $n$ input fields $A_i(\omega)$ and the qubit system being in a state labeled by $j$, the output in frequency space is
\begin{align}
Z_j(\omega)=\sum_{i=1}^n H_{j,i}(\omega) A_i(\omega),
\end{align}
where $H_{j,i}$ are the effective transfer functions of the light going through $\mathcal{Q}$, depending nonlinearly on the qubit states and on the frequency. In the dispersive regime these functions are diagonal in the qubit basis, i.e. the amplitudes $\hat{a}$ of the intracavity field couple to the $j\text{th}$ qubit via the Hamiltonian $\hat H_d=\chi_j\hat{a}^\dag\hat{a}\hat{\sigma}_j^{(z)}$. This represents a quantum non-demolition measurement and for timescales short compared to the qubits lifetimes we can neglect their dissipation and associated noise. For our purposes, we assume that the light pumping the system stems from coherent drives. Thus the quantum noise of the light is white vacuum noise and we can neglect noise operators in our notation. In this simplification, all functions can be considered to be the expectation value of coherent optical states, though this is not crucial to our derivation. 

As a typical example, a qubit-conditioned, single-port cavity has frequency-space dynamics governed by
\begin{eqnarray}
Z_j(\omega)&=&H^{(c)}_{j}(\omega) A(\omega)\nonumber\\
H^{(c)}_{j}(\omega) &=&\kappa_c/(i\Delta_c + i\chi_{j}-\frac{\kappa_c}{2}-i\omega)
,\label{eq:cav_transf}
\end{eqnarray}
with the sign of $\chi_j$ depending on the qubit state. Here  $\kappa$ is the cavity linewidth and $\Delta$ is the detuning of the pump field's carrier frequency from the ``empty cavity''. Cascaded arrangements have the output of one cavity as the input to the next. Thus, for instance, for two sequentially coupled single-port cavities, the amplitude reaching the detector is
\begin{eqnarray}
Z_{j,k}(\omega)&=& H_{j,k}(\omega) A(\omega)=H^{(1)}_k(\omega)H^{(2)}_j(\omega)A(\omega) ,\label{eq:singlecav}
\end{eqnarray}
where each $H^{(i)}_j$ is of the form (\ref{eq:cav_transf}) for cavity $i$. 

Determining which transfer function has acted on an input field constitutes a measurement of the qubit state.  Because the transfer functions relate inversely to the frequency, the typical temporal extent for the output fields is of the order $\kappa^{-1}$ , even if the input field is only nonzero for much shorter times.

This is illustrated in Fig.~\ref{fig:1Q} (a) and (b), which shows the intracavity field evolution for a single-port cavity conditioned by a qubit, subjected to a Gaussian measurement pulse with width much smaller than $T$. After that time the cavity does not return to zero. Instead, one must wait several inverse cavity decay constants to achieve a reset of the measurement apparatus. This prevents the system to be ready for the next operation and limits the duty-cycle of an experiment or device. 

\section{Direct, smooth solution}

The aim is to tailor the input function knowing the possible states of the qubit system and their effect on the total system's transfer function. This allows for input functions that lead to outputs that are constrained in time irrespective of the state the qubit system finds itself in. 

We can express the solution in terms of a trial input function $\Omega(t)$ and a linear combination of its derivatives, similar to the DRAG expansion \cite{Motzoi2013}, which itself is a multi-transition version of counter-diabatic methods \cite{shortcuts-1}. For a single input port we write the input field as
\begin{align}
A(t)\equiv\Omega(t)+\sum_{l=1}^n c_l i^l d^l \Omega(t) /dt^l\label{eq:pulse},
\end{align}
with $n$ the number of distinct quantum states. 
The function $\Omega$ and its first $n-1$ derivatives should all vanish at $t=0$ and $t=T$. These are not restrictive conditions, as such functions form an infinite dimensional vector space.

Formally, the depopulation of any excitation from all of the $n$ measured cavity modes requires all network element populations to adiabatically follow the input pulse shape  (according to their equations of motion, e.g.~\eqref{eq:cav_transf}). This is equivalent to removing the dependence on the derivative of their own state.  For a single measured state, this can be accomplished by adding a single derivative term $c_1\neq0$ to the input pulse that exactly cancels the derivative in Schroedinger's equation.  This allows arbitrarily fast measurement.

On the other hand, for multiple measured states, the coefficient in front of the derivative for each state is different, and so the counter-diabaticity condition is different for each of the states. The key insight is to move to a super-adiabatic basis to obtain the $n$ degrees of freedom needed.  The counter-diabaticity condition can then be exactly solved as a system of algebraic equations for the $n$ states. The derivation is given in Appendices A and B.

\begin{figure}	\label{fig:3Q}
	\includegraphics[width=.43\textwidth]{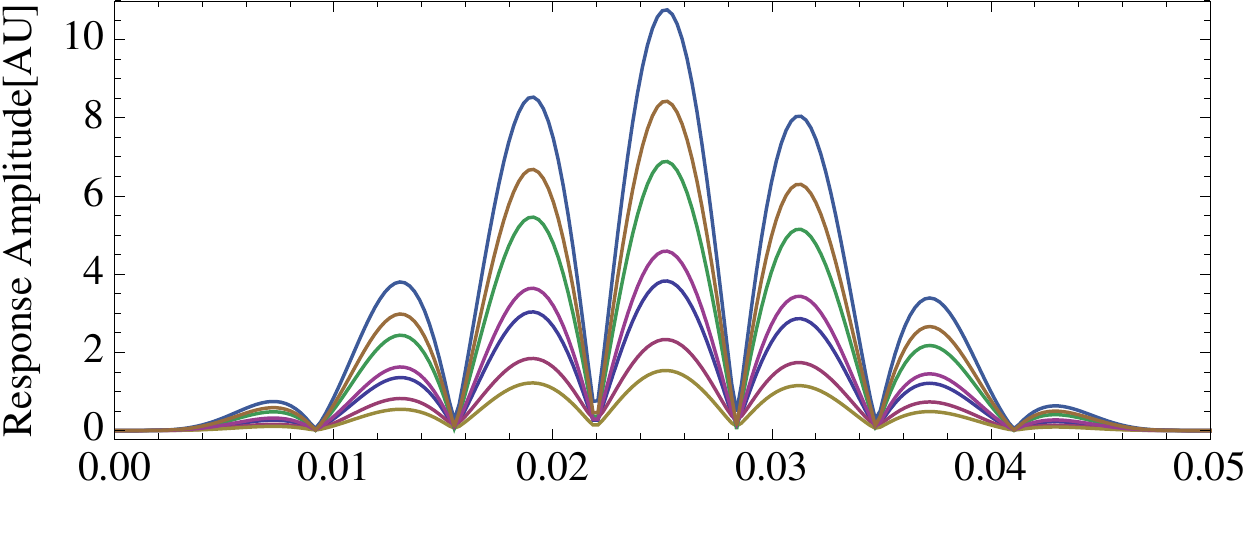}
	\includegraphics[width=.43\textwidth]{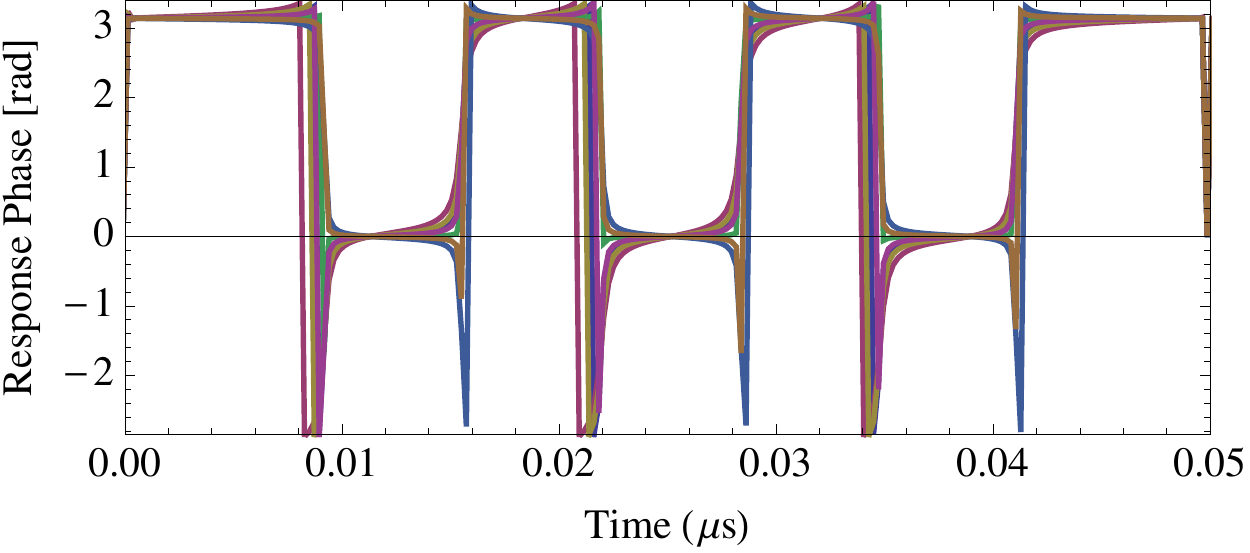}

\caption{Cavity evolution with 3 qubits with derivative corrections for a short measurement (50ns). Amplitude (top panel) and phase (bottom panel) of the 7 output signals (relative to the 000 state, c.f.~(7)) are plotted. We see from the two panels that most of the information is contained within a single quadrature. System parameters are the same as Fig.~2, with $\chi_1=3.6$MHz, $\chi_2=2.0$MHz, $\chi_3=1.1$MHz. }
\end{figure}

In this main text, we equivalently present a simpler, direct Ansatz solution in frequency space instead of solving the system of equations for its coefficients in the time-derivative basis. A direct solution for the same linear system can be obtained by  choosing the frequency dependence of the fields to be strictly polynomial in $\omega$. 
Consequently, the input field that unconditionally leads to short-timed field amplitudes in the system is 
\begin{eqnarray}
A(\omega)&\equiv&\prod_{k}H^{-1}_k(\omega)\Omega(\omega),\label{eq:fieldsolCD}
\end{eqnarray}
where $k$ runs over all measured states.  The $j$-th output signal then loses its inverse $\omega$ dependence and is given by
\begin{eqnarray}\label{outputzj}
Z_j(\omega)&=&\prod_{k\neq j}H^{-1}_k(\omega)\Omega(\omega).
\end{eqnarray}
The solution has the required property of losing its Lorentzian shape and depending only (linearly) on the derivatives of the trial function $\Omega(t)$. The intracavity and output fields are ensured to be nulled by the end of the measurement. It is straight-forward to calculate their evolution in time either via the inverse Fourier transform or by using  $\mathcal{F}(i^n d^n \Omega(t) /dt^n)=\omega^n \Omega(\omega)$. The procedure is easily generalized to multiple ports.

At any time $t$ during the measurement, the output field consists of a displacement independent of the system's state and an offset depending on the state of the quantum system that provides distinguishability of different quantum states,
\begin{align} 
{\bf \tilde Z}_j(t)={\bf \tilde Z}_0(t)+{\bf \tilde D}_j(t).
\label{eq:vectorJ} \end{align}
Bold faced quantities denote vectors in the $I-Q$ plane. The vector ${\bf \tilde Z}_0(t)$ is given by the terms in the inverse Fourier transform of Eq. (\ref{outputzj}) that do not depend on any $\chi_i$, while ${\bf \tilde D}_j(t)$ is the remainder. The efficiency of the measurement can thus be optimized to take full advantange of the fast measurement, as we do in Sec.~\ref{sec:dyne}.

Figs.~\ref{fig:1Q}(b), \ref{fig:1Q}(d) and \ref{fig:3Q} show the effect of the corrected measurement pulse in the quadrature plane for a single and three qubit system respectively. The cavity field exactly and smoothly returns to zero at the end of the pulse, as desired.  In both cases, the measurement signal stays within a single quadrature for the duration of the pulse, satisfying the the requirement of high efficiency.

\section{Pulsed light in quantum networks}

Multi-qubit measurements of non-trivial quantum networks \cite{Combes2017} offer a promising avenue for developing technologies involving entanglement generation \cite{Roch2014,Dickel2018}, non-local reservoir engineering \cite{Clark2003,RemoteBath} , scalable error correction \cite{Kerckhoff2009}, and metrology \cite{Komar:2014dp}.  The transfer function formalism is especially well suited for this setting because of the complex nature of the temporal dynamics in such open systems. We provide two prevalent examples.

\begin{figure}	
	\includegraphics[width=.35\textwidth]{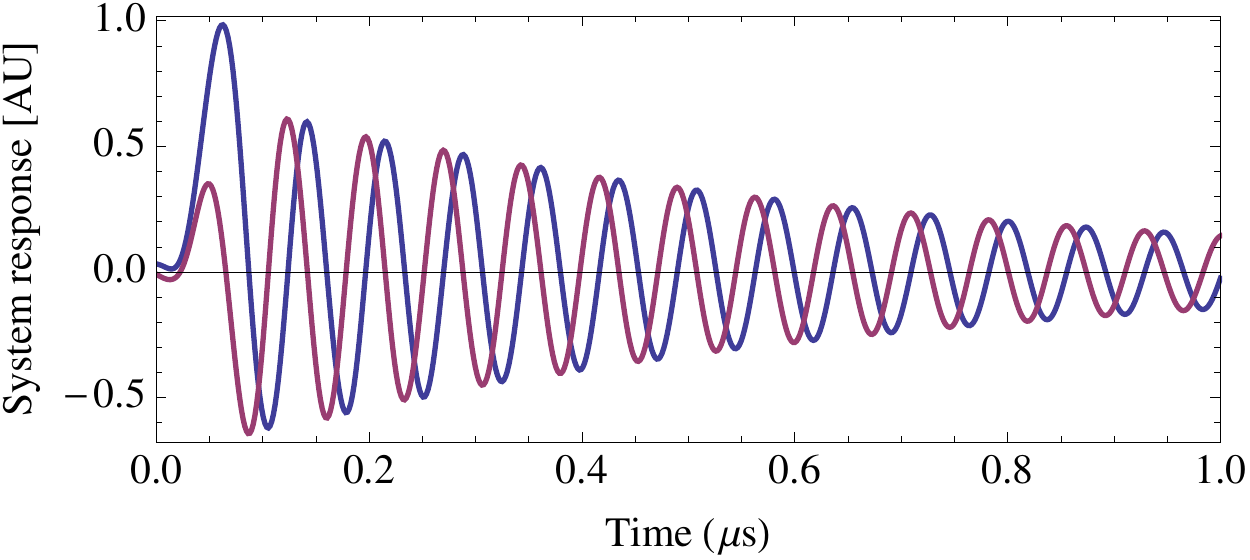}
	\includegraphics[width=.35\textwidth]{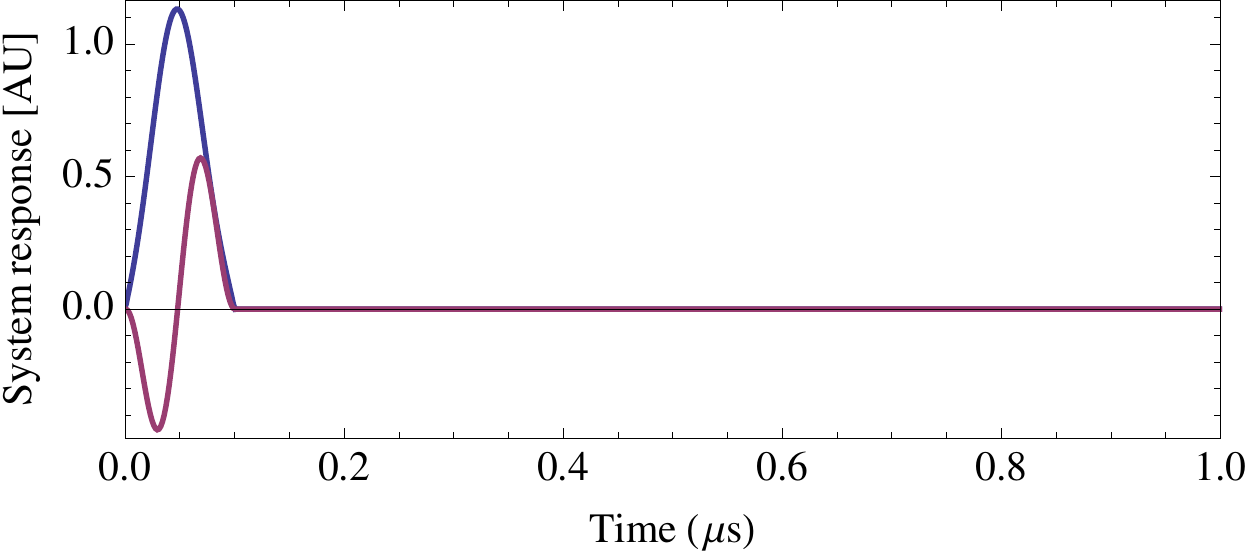}
\caption{Scaled output from a Purcell readout cavity measuring a single qubit, as described in the main text and Fig.~1(b). (a) Uncorrected Gaussian pulse readout. (b) Output when the same Gaussian pulse is augmented with derivative correction. The responses shown are for the qubit in the 0 state, but the 1 state behaves analogously. The system parameters are: inter-cavity coupling $G=20$MHz, detunings $\Delta=0.2$MHz and $\delta$=20MHz, dispersive shift $\chi=2$MHz, linewidth $\kappa=2$MHz. }\label{fig:Purcell}
\end{figure}
\subsection{Purcell filter}

A Purcell filter is used to tailor the noise of measurement cavities and extends coherence times in superconducting qubits \cite{Sete15}.  The network arrangement consists of two coupled cavities with input and output ports of the measurement field occurring on single cavities (Fig.~1(b)). For one qubit, the two intra-cavity fields \cite{Sete15} can be shown to obey 
\begin{eqnarray}
C_1(\omega) &=&(A(\omega)+G C_2(\omega))/(i\Delta + i\chi_{j}-i\omega)\\
C_2(\omega) &=&(G^* C_1(\omega))/(i\delta -\frac{\kappa}{2}-i\omega)
.\nonumber\label{eq:cavtransf}
\end{eqnarray}
The counter-diabatic, fourth derivative solution to this problem is thus
\begin{eqnarray}
A(\omega) &=&\Omega(\omega)H^{-1}_{0}H^{-1}_{1}\\
H^{-1}_{j}&=&(i\Delta + i\chi_{j}-i\omega)(i\delta -\frac{\kappa}{2}-i\omega)+|G|^2.\nonumber
\label{eq:cavtransf}
\end{eqnarray}
Note that while there are only 3 modes in the uncoupled cavities (as would occur for a cascaded system), because the cavities are directly coupled their modes hybridize and there are in fact 4 modes to cancel. Thus we require 4 rather than 3 derivatives to solve the adiabaticity condition. Not also that for G and/or $\chi$ small this can be reduced to 2 or 3 derivatives for approximate solutions. The simulation of measurement with a Purcell filter with and without shortcut are shown in Fig.~\ref{fig:Purcell}, reducing measurement times by up to more than an order of magnitude.

\subsection{Remote entanglement generation}
Measuring two cavities containing a qubit each can lead to the entanglement of the two remote qubits if different qubit states yield the same measurement, i.e. have the same effect on a continuous input field \cite{Roch2014,Dickel2018}. Postselection on the measurement result gives the desired entangled pair of remote qubits. 

The network of such a system gives the equations
\begin{eqnarray}
Z_{j,k}(\omega)&=& H^{(2)}_k(\omega)H^{(1)}_j(\omega) A(\omega)+  H^{(2)}_k(\omega) B(\omega)\quad\quad \label{eq:cascadeentang}
\end{eqnarray}
where $B$ is an input on a weakly coupled backport to the second sequential cavity (Fig.~1(c)).

The indistinguishability needed for entanglement can be obtained by using $B$ to compensate for the difference in outputs given input $A$, arising from static parameter differences between the two cavities \cite{Motzoi2015}. The  condition for indistinguishability of arbitrary states $01$ and $10$ is \begin{eqnarray}
Z_{0,1}(\omega)&=&Z_{1,0}(\omega).\label{eq:paritycond}
\end{eqnarray}
This allows to solve for the $B$ field that satisfies the constraint. However, the known solution (Appendix C) will result in a pulse sequence that does not minimize the measurement time, leading to preparation of states with low coherence. To overcome this we may apply the measurement fields
\begin{eqnarray}
A(\omega)&=&\prod_{i}\prod_{k}\frac{\Omega(\omega)}{H^{(1)}_i (\omega) H^{(2)}_k(\omega)},\quad B(\omega)=\prod_{k}\frac{\Gamma(\omega)}{H^{(2)}_k(\omega)},\nonumber
\end{eqnarray}
where $\Gamma$ is a trial function for the weak input on the second cavity. 
Plugging this into \eqref{eq:cascadeentang} along with the indistinguishability criteria \eqref{eq:paritycond} simplifies to
\begin{eqnarray}
\Gamma(\omega)&=&\frac{H^{(2)}_0(\omega)/ H^{(1)}_0(\omega)-H^{(2)}_1(\omega)/ H^{(1)}_1(\omega)}{H^{(2)}_1(\omega)-H^{(2)}_0(\omega)}\Omega(\omega)\nonumber\label{eq:parityCD}
\end{eqnarray}
For the specific transfer function \eqref{eq:cav_transf}, the solution can be further simplified, as shown in Appendix C.

\begin{figure}	
	\includegraphics[width=.45\textwidth]{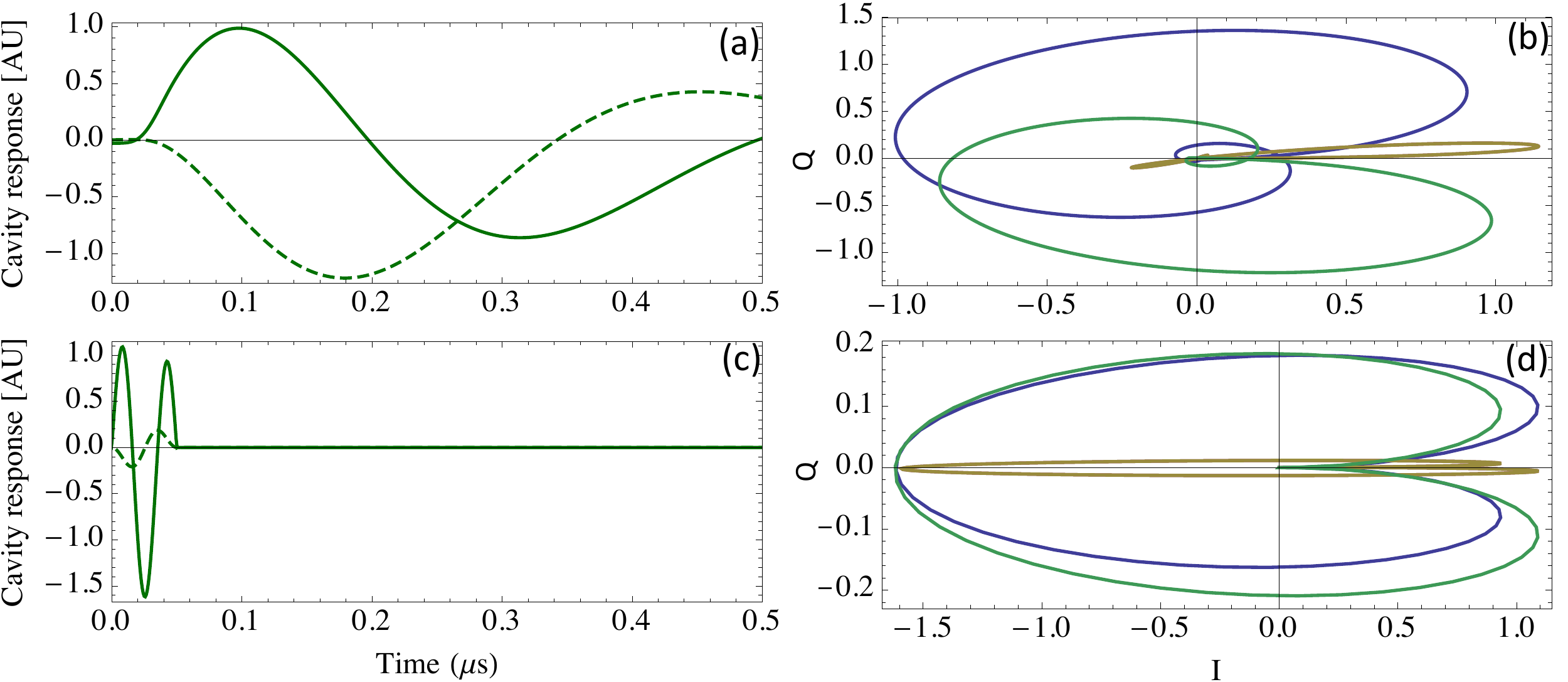}
\caption{Cascaded cavity setup for entanglement generation as in Fig.~1(c). Top row is for a Gaussian measurement pulse and bottom row is the same pulse with derivative corrections.  (a) and (c) show the response of the cavity when the qubits are in the 11 state, with solid and dashed lines corresponding to I and Q quadratures, respectively.  (b) and (d) show the respective evolutions in the I-Q plane for all four qubit states. Note that corrective pulses on the second cavity have been applied in both cases to ensure indistinguishability within the single excitation subspace (01 and 10). }\label{fig:2Q}
\end{figure}

The resultant cavity dynamics are shown in Fig.\ref{fig:2Q} with all cavity fields returning to zero at much shorter times, and homodyne detection rendered possible.

As an extension of these techniques, one can also apply feedback on the results of the quantum network, e.g. for error correction or to obtain deterministic entanglement generation.  For instance, the applied feedback fields $A$ and $B$ can be conditioned on the measured signals.  To adiabatically follow the trial functions during the process, one can apply the same counter-diabatic procedure, e.g.~\eqref{eq:paritycond}, with the time dependence of the feedback trial functions implicitly through the measurement current $\xi(t)$: $\Omega(t)=\tilde\Omega(\xi(t))$ and $\Gamma(t)=\tilde\Gamma(\xi(t))$.

\section{Common-quadrature measurement}\label{sec:dyne}

The performance of quantum measurement is optimal when all information is obtained through commuting observables. In the context of a field measurement, this means measuring only a single quadrature of the output field at any given time. Technically, this is achieved by mixing the output field with a coherent field with the same carrier frequency to achieve balanced homodyne detection.  The phase angle of the measured quadrature may also vary with time in a synodyne measurement known from cavity optomechanics \cite{Buchmann2016}

A heterodyne measurement of the output signal will reveal the entire trajectory in the $I-Q$ plane, however, it does so at the expense of additional noise and, in the case of circuit QED at the expense of additional equipment on and off-chip. We consider two measurement protocols that do not add any superflous noise: homodyne and synodyne. A homodyne measurement yields $c_j^\textrm{(hom)}(t,\alpha)={\bf e}_\alpha\cdot{\bf \tilde Z}_j(t)$, where ${\bf e}_\alpha$ unit vector with polar angle $\alpha$ in the $I-Q$ plane. For a synodyne measurement this angle is additionally a function of time whose evolution may be chosen freely and we denote the corresponding measurement traces $c_j^\textrm{(syn)}(t,\alpha(t))$.  Note that an optimal synodyne measurement will always outperform the homodyne case, but often the homodyne case may already be near-optimal.

The measurement traces can be visualized as the projection of the curve ${\bf \tilde Z}_j(t)$ onto a line in with polar angle $\alpha$ in the $I-Q$ plane.  The distinguishability of two states $j$ and $j'$ can be quantified by the integral of the absolute value of their difference integrated over the measurement duration $T$
\begin{align}
\mathcal{Q}_{j,j'}^\mu(\alpha)=\int_0^T dt|c_j^\mu(t,\alpha(t))-c_{j'}^\mu(t,\alpha(t))|,
\end{align}
where $\mu$ stands for homodyne or synodyne respectively. 
The larger the value of $\mathcal{Q}_{j,j'}$ is, the easier the applied pulse distinguishes between the states $j$ and $j'$.

For instance, for a single qubit we have 
\begin{eqnarray}
D^{qubit}_\pm(\omega)&=&i\chi_{\mp}\Omega(\omega)/\kappa.
\end{eqnarray}
The offset has the same complex phase for both states and optimal distinguishability is possible in a homodyne measurement. From the evolution of the fields in Fig.~2 it is clear that projecting the two curves onto the line of symmetry between the two curves gives all information distinguishing the two states.

If the system has more than two states available, it is not possible to distinguish between all states with a homodyne measurement with optimal efficiency, though near-optimal performance is not out of the question. In the already discussed examples, we see the phase relation between states evolving in time.  For a three-qubit system, we see in Fig.~\ref{fig:3Q} that the vectors in time are all well lined up for fast measurement. The phase does evolve but it remains mostly parallel or anti-parallel for most of the measurement and so we can conclude for these typical parameters that synodyne measurement will give a small boost in measurement contrast, but is by no means necessary. This implies that the local oscillator does not need to be shaped.

\begin{figure}\label{fig:dyne}
\includegraphics[width=0.4\textwidth]{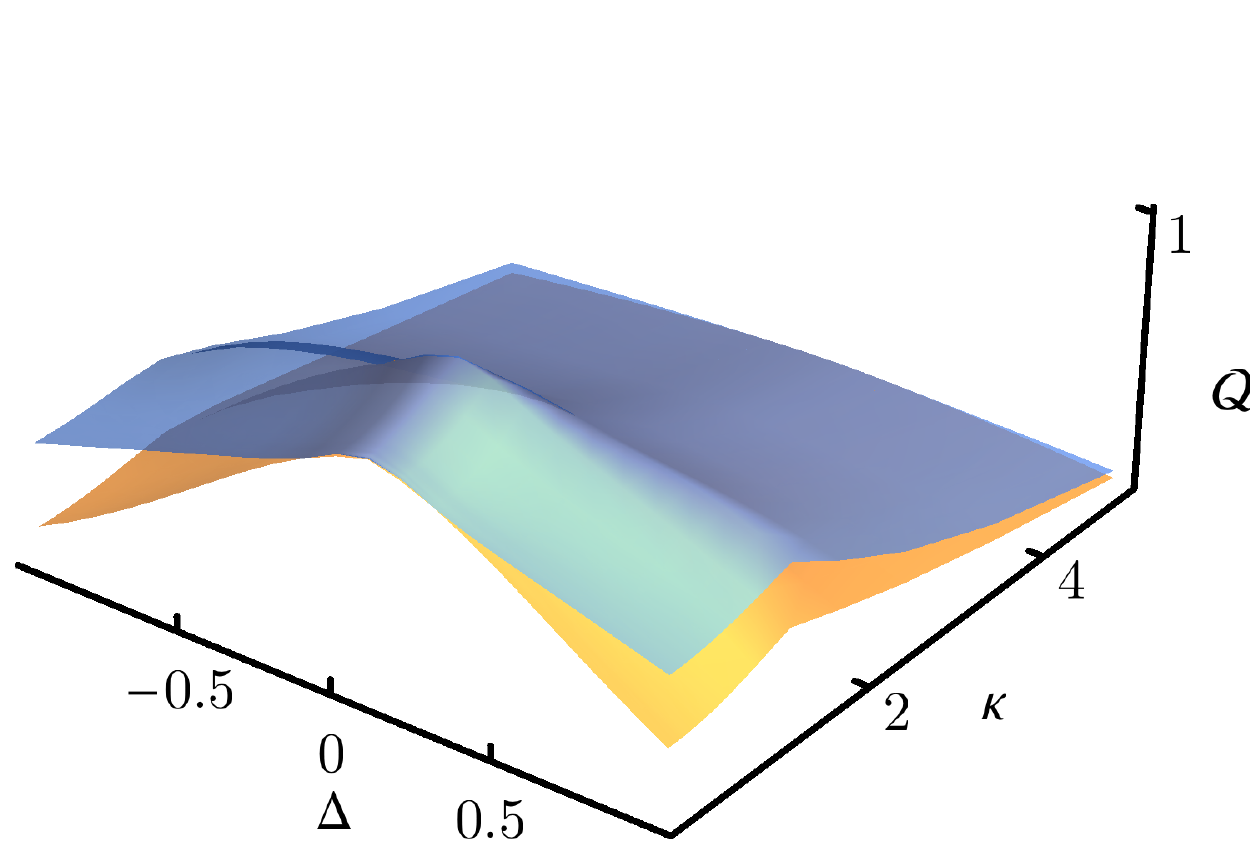}
\includegraphics[width=0.4\textwidth]{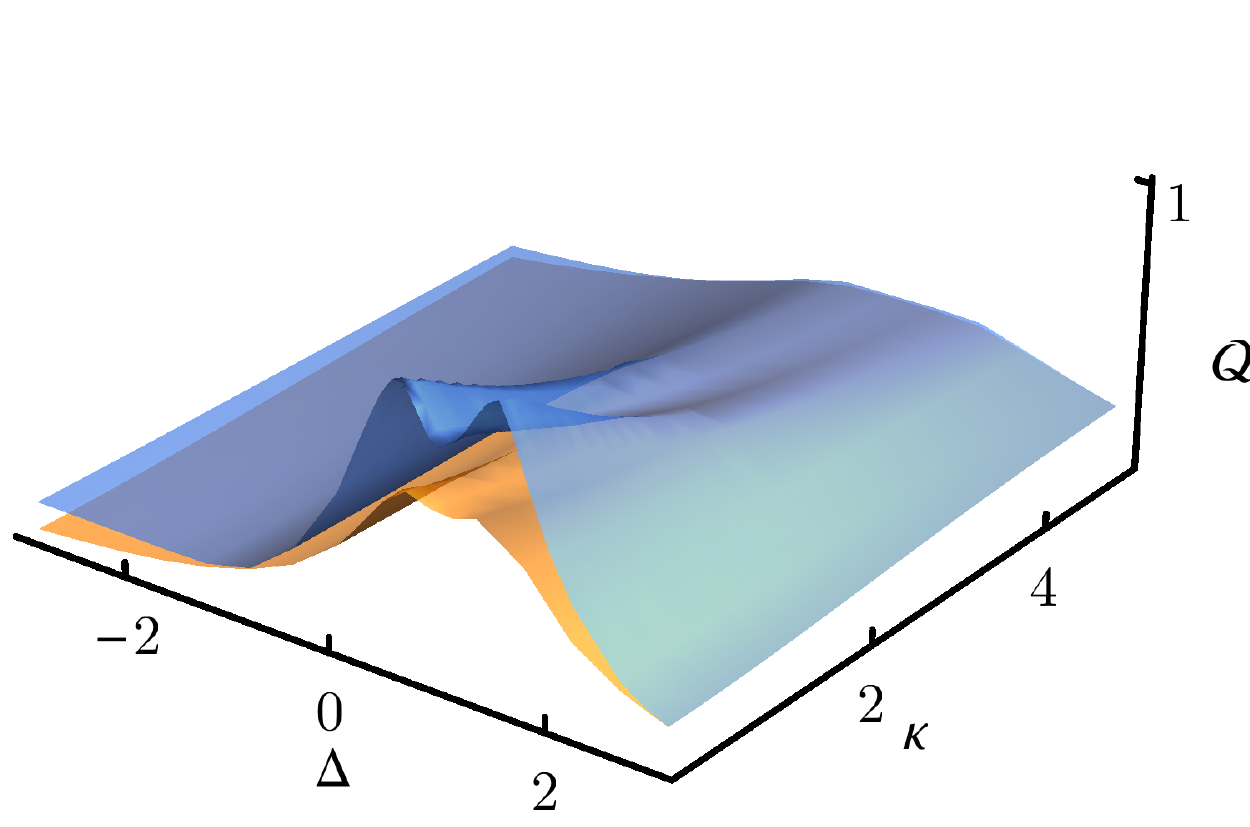}
\caption{Integrated distinguishability of the two least distinguishable states in the measurement of a four-state system. The top surface is a synodyne measurement while the one below is the homodyne measurement. Distinguishability is normalized to the maximum distinguishability of the synodyne protocol. System parameters are $T=1$ for both figures and $\chi_{1,2}=\pm 1, \chi_3 = 3$ for panel (a) and $\chi_{1,2}=\pm 0.1, \chi_3=0.3$ for panel (b). }
\end{figure}

More systematically, we can compute the optimal measurement angle $\alpha$ for different parameter ranges 
by maximizing the distinguishability at every time. 
For the least distinguishable states $j$ and $j'$, the optimal homodyne angle choice will be given by 
\begin{align}\label{eq:distinghom}
\max_{\alpha\in[0,2\pi)}\min_{i,j}\left[\mathcal{Q}_{i,j}^{\text{(hom)}}(\alpha)\right].
\end{align}
In the case of a synodyne measurement, we can find the ideal time-dependent phase angle for every instance in time, by using only the state-dependent component of the signal \eqref{outputzj}, as in
\begin{align}\label{eq:disting}
\max_{\alpha\in[0,2\pi)}\min_{i,j}\left[({\bf D}_i(t)-{\bf D}_j(t))\cdot{\bf e}_\alpha\right].
\end{align}
The result is a time-dependent homodyne angle that can be adapted via a tuning the phase of the local oscillator of a homodyne measurement. 
We give a general sense of the distinguishability using homodyne and synodyne measurements for fast and slow measurement relative to the dispersive shift energy, in Fig.~\ref{fig:dyne}A and \ref{fig:dyne}B respectively. In this case, we treat a 3-level system and we vary detuning and cavity linewidth.

We limit the maximal intracavity field amplitude to normalize the comparison (assuming this is the limiting factor, due to breakdown of the dispersive approximation). We find that if the dispersive shifts are large compared to $1/T$, the ideal distinguishability attains a maximum at $\kappa T\simeq 1$ and $\Delta$ close to $0$. For such parameters, the syndoyne procedure does not yield significant benefits over a single quadrature measurement. This changes considerably if the dispersive shifts are smaller than $1/T$. In that case, the ideal distinguishability increases with decreasing $\kappa$  and a synodyne measurement outperforms a homodyne measurement by almost a factor of $2$. 

This exploration of parameters matches the intuition from Fig.~\ref{fig:3Q} for the eight-state system, whereby for fast pulses the phase evolution is mostly dictated by the dispersive shifts multiplied by the time-dependence of the pulse (being still adiabatic).  Thus for short times we see the added benefit that not only does the decoherence in the system greatly decrease but the efficiency of the measurement can also be higher.

\section{Conclusion}

We have demonstrated a simple, smooth pulse sequence protocol that dramatically reduces dispersive measurement times using readout cavities and quantum networks.  Compared to digital-pulse numerical sequences, the speed for high-fidelity measurement operations is reduced considerably, even when experimental non-idealities are considered.  The analytical closed-form solutions are naturally robust to noise and low-bandwidth filtering and do not add any calculational or hardware overhead. For well calibrated systems, the pulses should work out-of-the-box without the need for experimental closed-loop optimization. 

The shaping of measurement pulses demonstrates that readout need not be the limiting factor in computing clock cycles as is currently the case. Since coherent measurement light couples relatively easily in and out of quantum systems, the quantum speed limit of universal computation is more likely bottlenecked by two-qubit gates given currently limited interaction strengths. 

Our scheme transcends single qubits and is applicable to complex quantum networks containing a multitude of states and cavities, while the simplicity, robustness, and high-speed are preserved.  This directly impacts current headway in using Purcell filters to shield qubits from the environment, in measuring and counter-acting leakage in three-level systems while retaining homodyne measurement efficiency, and in scaling up to larger systems. These aspects make the demonstrated techniques particularly useful for efficient quantum error correction.

Although we have limited our analysis to dispersive readout schemes, the transfer function formalism also applies to quantum networks involving single photons. Thus, all the methods described can also be used to speed up entry and exit of pulses into non-dispersive cavities \cite{Motzoi2018,Cohen2018}. The implications for such situations will be subject of further investigations. 

\newpage

\bibliographystyle{apsrev4-1}
\bibliography{simple_smooth_fast}

\appendix

\section{general counter-diabatic and super-adiabatic methodology}

The adiabatic theorem states that if a system is initially prepared in some instantaneous eigenstate $\ket{e_n(t)}$ of its time-dependent Hamiltonian $\op{H}(t)$, the state evolved according to Schr{\"o}dinger's equation
$	
i\partial_t \ket{\psi(t)}  = \op{H}(t)\ket{\psi(t)}, 
$
will follow the instantaneous eigenstate, i.e. $\ket{\psi(t)}  =\ket{e_n(t)}$, up to global phase, provided the Hamiltonian $\op{H}(t)$ changes sufficiently slowly \cite{adiabatic-theorem-4}. The diabatic error will exactly equal
\begin{align}
I=\sum_n\sum_{m\neq n}\frac{\bra{e_n(t)}\partial_t \op{H}(t)\ket{e_m(t)}}{E_n-E_m}\ket{e_m(t)}\bra{e_n(t)}\label{eq:diaberr}
\end{align}
Equivalently, in the adiabatic eigenbasis of $\op{H}_0(t)$ (that is the instantaneously diagonalized basis), via a unitary transformation $\op{V}_0(t)=\sum_n\ket{n}\bra{e_n(t)}$, the system Hamiltonian is given by
\begin{align} \label{eq:adiabatic_basis}
  \op{H}_{\rm eff}(t) & =  \op{V}_0(t) \op{H}_0(t) \op{V}_0(t)\dg + i \dot{\op{V}}_0(t) \op{V}_0(t)\dg.
\end{align}
To make the adiabatic eigenbasis exact, a well-known technique is to cancel the inertial term, $I=i \dot{\op{V}}_0(t) \op{V}_0(t)\dg$, equal also up to a geometrical phase to (A1), with an additional control term in the Hamiltonian.
\begin{align}
  \op{H}_{\rm cd}(t)=-I, 
  \end{align}
Hence, time evolution of Hamiltonian $H+H_{CD}=\op{V}_0(t) \op{H}_0(t) \op{V}_0(t)\dg$ can be carried out analytically to  equal
\begin{align}\label{eq:propagator_diag-2}
	\op{U}(t) & = \sum\limits_{n} \exp{-i\int\limits_0^t E_n(t')\, t'}\ketbra{e_n(t)}{e_n(0)}.
\end{align}
 
A useful complementary framework is that super-adiabaticity\cite{superadiabaticity-berry-1,superadiabaticity-berry-2}, also formulated 20 years earlier by Garrido\cite{superadiabaticity-2} in the context of adiabatic invariants. The key idea is to utilize a sequence of iterative adiabatic transformations to account for finite inertial terms $I$ arising at each iteration. That is, analogously to (A2), we define the $j$-th adiabatic frame as
\begin{align}\label{eq:superadiabaticity_1}
	\op{H}_{j} & = \op{V}_{j-1} \op{H}_{j-1} \op{V}_{j-1}\dg + i\dot{\op{V}}_{j-1} \op{V}_{j-1}\dg, \quad j \geq 1,
\end{align}
where all operators are in general time-dependent. The instantaneously diagonal and the inertial terms are given by $\op{D}_j \equiv \op{V}_{j-1} \op{H}_{j-1} \op{V}_{j-1}\dg$ and $\op{I}_j \equiv i(\partial_t \op{V}_{j-1}) \op{V}_{j-1}\dg$, respectively. 

Canceling $I_j$ in the given frame allows for cancellation of the transition leakage.  In fact, any combination of the $j$ frames can be used to cancel the diabatic error, i.e.
\begin{align}\label{eq:CD_Ham_iterative}
	\op{H}_{\rm cd}^{(j)} & = -i\sum\limits_{k=1}^{n}a_k (\prod\limits_{l=1}^{k-1}V_l)\dg \dot{\op{V}}_j (\prod\limits_{l=1}^{k}V_l),\quad \sum a_k = 1.
\end{align}
These different terms typically span the temporal basis of the control and the basis can be used in principle to remove not one but $n$ unwanted transitions.  In the next section we show how this can be done exactly for measurement via resonators on $n$ different atom-cavity hybridized states.

\section{exact solution to counter-diabaticity criteria for multiple unwanted cavity mode excitations}

To obtain a measurement signal that starts and stops at zero for a short duration, we solve for the sufficient condition that the frequency dependence of the cavity is only polynomial in $\omega$. We use the basis to define our input pulses in terms of the derivatives of a trial function (here Gaussian pulse) defined so that all relevant derivatives are zero at the temporal boundaries of the pulse. Thus, in the time domain,
\begin{align}
A(t)=\Omega(t)+\sum_n b_n i^n d^n \Omega(t) /dt^n. \label{eq:basis}
\end{align}
The equations of motion of the simple single-cavity system are
\begin{align}
\{(-\partial_t+E_l) C(t)=\Omega(t)+\sum_n(\partial_t)^n b_n \Omega(t)) \} \label{eq:DEcav}
\end{align}
where the complex energies are $E_l\equiv  i\Delta+i\chi_j-\frac{\kappa}{2}$ for state $l$ and include cavity relaxation.

The adiabatic diagonalization of the measurement field is given by the displacement operator. The (first) adiabatic frame is given by the displacement 
\begin{align}
V_0(t)=D(\alpha_0)=D(\Omega/E_l)=\exp (\alpha_0^* a-\alpha_0 a\dg).
\end{align}
The first inertial term is given by $I_1=\dot V_0 V_0\dg= \partial_t(i \alpha_0^* a - i \alpha_0 a\dg)$.  Each subsequent frame iteration is similarly given by 
\begin{align}
V_j&=D(\alpha_j)=D(i\dot\alpha_{j-1}/E_l)\\
I_j&=\partial_t(i \alpha_j a - i \alpha_j^* a\dg).
\end{align}
Note that each state measured will have a different $E_l$ and therefore the  transformations will need to be different for each state

We choose the $b_j$ to partially counteract the inertial terms at each iteration.  The consecutive displacements of the cavity correspond to superadiabatic iterations.  The $j-th$ counter-diabatic term removes $b_j I_j$ from the cavity field. In the subsequent superadiabatic iteration the inertial field would be $(1-b_j) I_{j+1}$ and the counter-diabatic term would be  $b_{j+1} (1-b_j) I_{j+1}$, and so forth for higher iterations.

We can expand the cavity field (like the measurement field) in the derivative basis: $C_l(t)=C^{(0)}_l(t)$=$\sum_{n=0}^N c_{l,n} i^n d^n \Omega(t) /dt^n$. Each superadiabatic iteration displaces to the vacuum the component with lowest order, i.e. $C_l^{(j)}(t)=\sum_{n=j}^N c_{l,n} i^n d^n \Omega(t) /dt^n$, where we discount scalar terms. Thus the $N$-th superadiabatic state will just be the vacuum, $C_l^{(N)}(t)=0$.

It can be verified by plugging into (B2) the coefficients
\begin{align}
b_1&=\frac{1}{E_1}+\frac{1}{E_2}+\cdots+\frac{1}{E_N}\nonumber\\
b_2&=\frac{1}{E_1 E_2}+\frac{1}{E_1 E_3}+\frac{1}{E_2 E_3}+\cdots+\frac{1}{E_{N-1} E_N}\nonumber\\
b_2&=\frac{1}{E_1 E_2 E_3}+\frac{1}{E_1 E_2 E_4}+\frac{1}{E_2 E_3 E_4}+\cdots+\frac{1}{E_{N-2} E_{N-1} E_N}\nonumber\\
\vdots&\nonumber\\
b_j&= \frac{1}{j!}\sum_{s_1=1}^N\sum_{s_2\neq s_1}\sum_{s_f\neq s_1,s_2}\cdots\sum_{s_j\neq s_1,\cdots ,s_{j-1}}\frac{1}{E_{s_1} E_{s_2}E_{s_3} \cdots E_{s_j} }
\end{align}
that the cavity superadiabatic coefficients are given by
\begin{align}
c_{l,j}&= \frac{1}{j!}\sum_{s_1\neq l}\sum_{s_2\neq l,s_1}\cdots\sum_{s_j\neq l,s_1,\cdots ,s_{j-1}}\frac{1}{E_lE_{s_1} E_{s_2} \cdots E_{s_j} }
\end{align}
and that the superadiabatic expansion indeed converges to the vacuum.
Because the field solution is an $(N-1)$-th order polynomial in $\partial_t$, it will start and end at zero if the first $N-1$ derivatives of $\Omega$ also start and end at zero. This solution is thus exact and smooth and does not have any intrinsic quantum speed limit based on system parameters.

\section{Indistinguishabiity criteria for entanglement of cascaded systems}

 The network for a cascaded two cavity system is given by the system of equations
\begin{eqnarray}
Z_{j,k}(\omega)&=& H^{(2)}_k(\omega)H^{(1)}_j(\omega) A(\omega)+  H^{(2)}_k(\omega) B(\omega)\quad\quad\quad\label{eq:cascade_entang}
\end{eqnarray}
where $B$ is an input on a weakly coupled backport to the second sequential cavity.
Applying the correction scheme for time-constrained cavities gives
\begin{eqnarray}
Z_{i,k}(\omega)&=&\prod_{j\neq i}\prod_{l\neq k}\frac{\Omega(\omega)}{H^{(1)}_j (\omega)H^{(2)}_l(\omega)}+\prod_{l\neq k}\frac{\Gamma(\omega)}{ H^{(2)}_l(\omega)}\label{eq:parityCD}\\
A(\omega)&=&\prod_{i}\prod_{k}\frac{\Omega(\omega)}{H^{(1)}_i (\omega) H^{(2)}_k(\omega)},\quad B(\omega)=\prod_{k}\frac{\Gamma(\omega)}{H^{(2)}_k(\omega)},\nonumber
\end{eqnarray}
where $\Gamma$ is a trial function for the weak input on the second cavity. 
The indistinguishability needed for entanglement can be obtained by using the back port (B) of the second cavity to compensate for the difference arising from uncontrolled (static) system parameters on input A, as in \eqref{eq:cascade_entang}. The  condition for indistinguishability of some arbitrary states $01$ and $10$ is given by
\begin{eqnarray}
Z_{0,1}(\omega)&=&Z_{1,0}(\omega),\, \text{or}\\
B(\omega)&=&\frac{ H^{(2)}_0(\omega)  H^{(1)}_1(\omega)- H^{(2)}_1(\omega)  H^{(1)}_0(\omega)}{H^{(2)}_1(\omega)-H^{(2)}_0(\omega)}A(\omega)\nonumber\label{eq:parity2Q}
\end{eqnarray}
where $j,k\in\{0,1\}$.  However, this known solution will result generally in a pulse sequence that does not minimize the measurement time. To overcome this, we instead apply the counter-diabatic fields \eqref{eq:parityCD}, which similarly simplify to
\begin{eqnarray}
\Gamma(\omega)&=&\frac{H^{(2)}_0(\omega)/ H^{(1)}_0(\omega)-H^{(2)}_1(\omega)/ H^{(1)}_1(\omega)}{H^{(2)}_1(\omega)-H^{(2)}_0(\omega)}\Omega(\omega)\quad\quad\quad\label{eq:parity2QgenCD}
\end{eqnarray}
For the specific transfer function \eqref{eq:cavtransf}, the compensation pulse will then take the form
\begin{eqnarray}
\Gamma(t)&=&\left(\frac{\chi^{(2)}_0\chi^{(1)}_1-\chi^{(2)}_1\chi^{(1)}_0}{\chi^{(2)}_1-\chi^{(2)}_0}+(\Delta_1+i\kappa_1/2)\right)\Omega(t)\nonumber\\
&&+\frac{(\Delta_2+i\kappa_2/2)(\chi^{(1)}_1-\chi^{(0)}_1)}{\chi^{(2)}_1-\chi^{(2)}_0}\Omega(t)\nonumber\\
&&+i\frac{\chi^{(2)}_0+\chi^{(1)}_1-\chi^{(2)}_1-\chi^{(1)}_0}{\chi^{(2)}_1-\chi^{(2)}_0}\dot\Omega(t)\quad\quad\quad\label{eq:parity2QCD}
\end{eqnarray}

\begin{figure}\label{fig:5}
\includegraphics[width=0.3\textwidth]{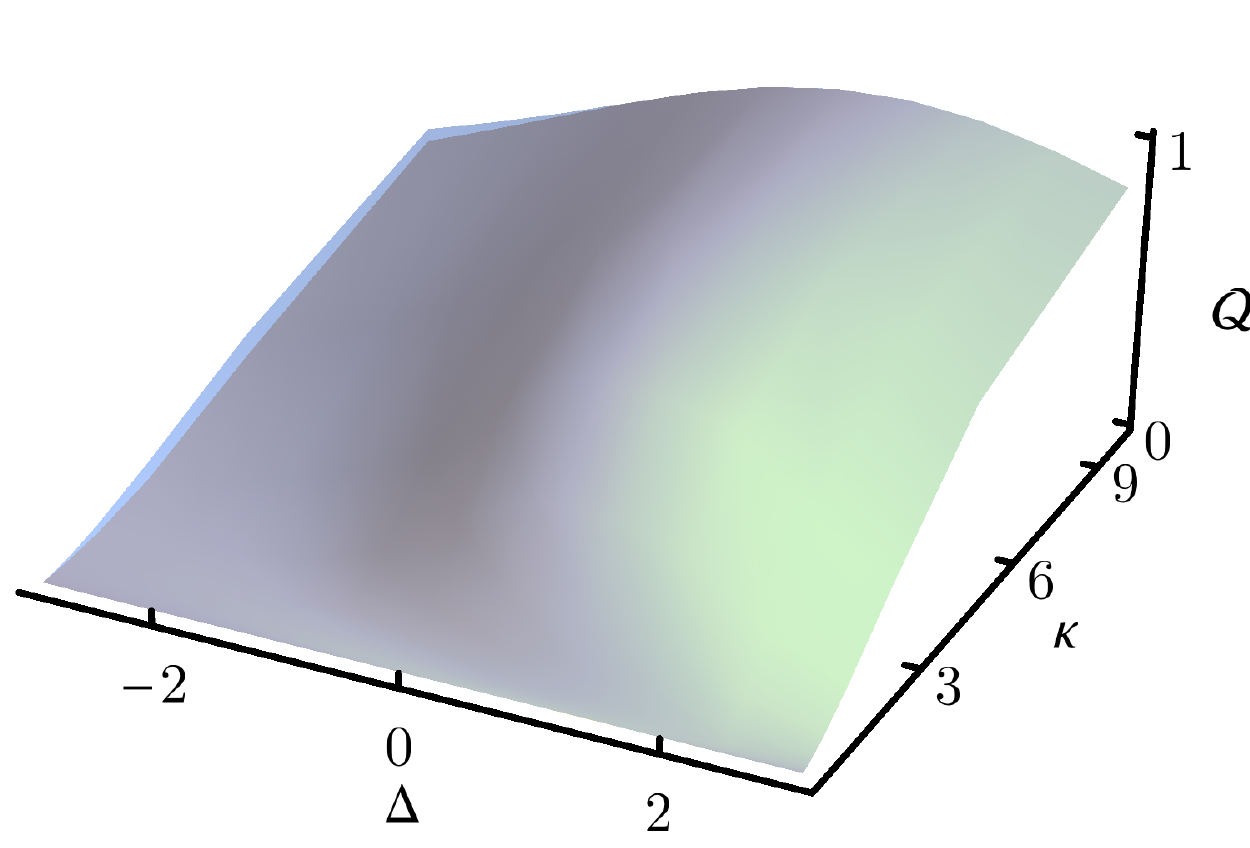}
\includegraphics[width=0.3\textwidth]{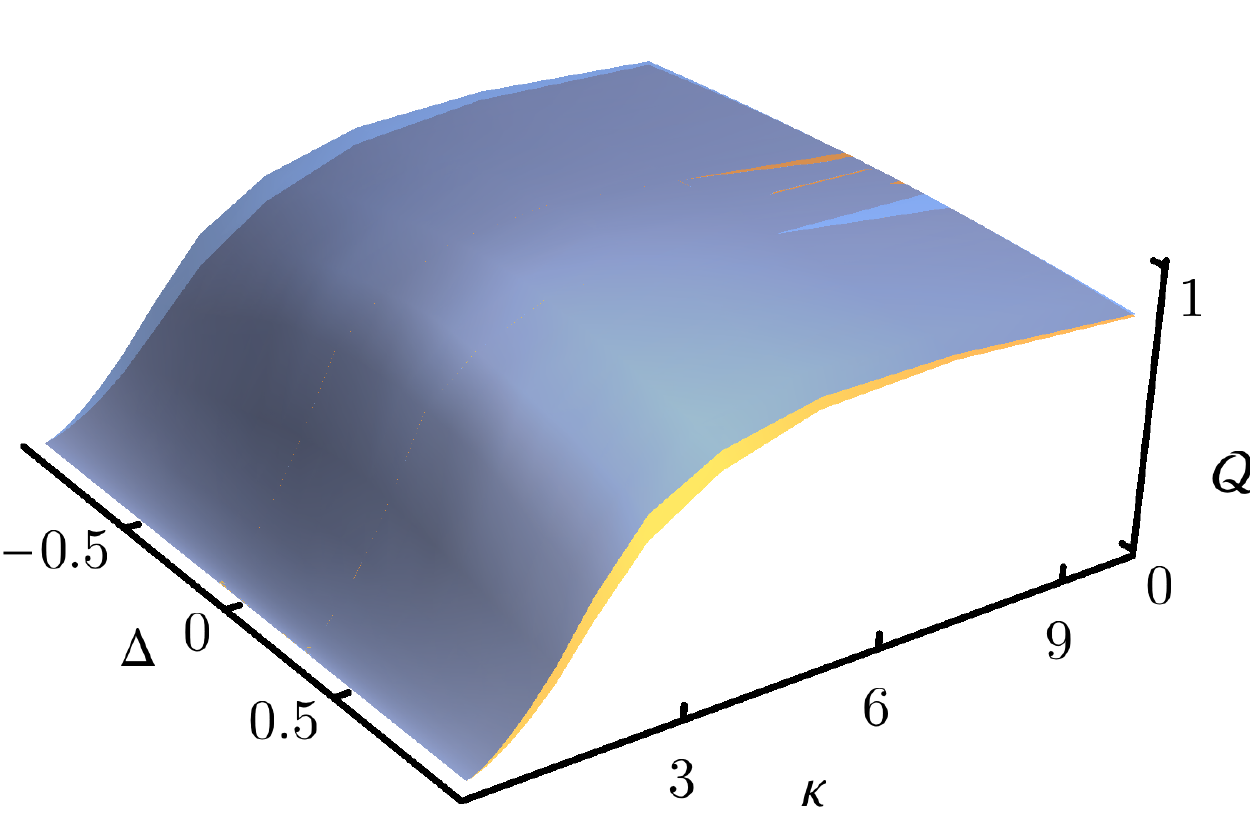}
\caption{Integrated distinguishability of the two least distinguishable states in the measurement of a three-state system with a constrained maximum power of the input fields. The top surface is a synodyne measurement while the one below is the homodyne measurement. Distinguishability is normalized to the maximum distinguishability of the synodyne protocol. System parameters are $T=1$ for both figures and $\chi_{1,2}=\pm 1, \chi_3 = 3$ for the top panel and $\chi_{1,2}=\pm 0.1, \chi_3=0.3$ for the bottom panel. }
\end{figure}

\section{AWG power limitated regime}

If instead of constraining the maximum field inside the cavity, we constrain the measurement field power, we can once again compare the ability to distinguish states for different parameter ranges.  We once again consider a three level system and both homodyne and synodyne measurement.  In general then it is best to keep $\kappa$ small in all cases to maximize the intra-cavity field that measures the qutrit, as can be seen in Fig.~7.  The dependence on the detuning is still there, with a preference for small positive detuning.  Overall, there is not much difference between keeping the homodyne phase constant or using synodyne-type phase variation, perhaps reflecting the fact that photons stay only briefly inside the cavity.

\end{document}